# Indirect multiphoton scattering between light and bulk plasmons via ultrafast free electrons

Ruoyu Chen[1,2], Jun Li[3]*, Qiaofei Pan[4,5], Dingguo Zheng[3], Bin Zhang[6], Ye Tian[2], Jianqi Li[3], Huaixin Yang[3]*, Yiming Pan[1]*

[1] *State Key Laboratory of Quantum Functional Materials, School of Physical Science and Technology and Center for Transformative Science, ShanghaiTech University, Shanghai 200031, China*

[2] *State Key Laboratory of Ultra-intense Laser Science and Technology, Shanghai Institute of Optics and Fine Mechanics, Chinese Academy of Sciences, Shanghai, China*

[3] *Beijing National Laboratory for Condensed Matter Physics, Institute of Physics, Chinese Academy of Sciences, Beijing, China*

[4] *Institute of Precision Optical Engineering, School of Physics Science and Engineering, Tongji University, Shanghai, 200092, China*

[5] *MOE Key Laboratory of advanced micro-structure materials, Shanghai, 200092, China*

[6] *School of Electrical Engineering - Physical Electronics, Center of Laser-Matter Interaction, Tel Aviv University, Ramat Aviv 69978, Israel*

## Abstract

Efficient coupling between light and bulk plasmons (BPs) remains a central challenge because of their inherent mode mismatch, limited penetration depth, and pronounced resonant energy mismatch between visible-range photons and BPs. In this work, we demonstrate that ultrafast free electrons can coherently mediate an interaction between electromagnetic fields and BPs at nanoscale. An electron pulse emitted from the photocathode of ultrafast transmission electron microscope, functions as a quantum intermediary that is capable of simultaneously interacting with laser field by multiphoton processes, and BPs by perturbative scattering. Electron energy-loss spectroscopy can capture this indirect interaction: the final electron energy distribution encodes both quantum pathways arising from distinct combinations of multiphoton absorption/emission and BP scattering events. Interference among these pathways gives rise to characteristic spectral modulations, directly revealing the exchange of energy and information between photons and BPs via the electron delivery. Our results show that femtosecond-driven, ultrafast electrons provide a viable route to modulate and even control bulk plasmon excitations in a volume, thereby extending beyond the conventional nanoplasmonics schemes on manipulating surface plasmons by light. This indirect light–BP interaction paves the promising way for exploring fundamental light–matter interaction at ultrafast and nanometer scales.

Light–matter interaction at the nanoscale underpins the burgeoning field of nanoplasmonics that explore and engineer the optical properties of metallic nanostructures and their interactions with light[1,2]. Back in the 1950s, electron energy loss spectroscopy (EELS) unequivocally identified two collective electron density oscillations in metals[3]: surface plasmon polaritons (SPPs)[4–6]- confined to metal–dielectric interfaces - and bulk plasmons[5–7], which freely propagate within the material[5,6]. SPPs, by virtue of their ability to mediate strong coupling between optical fields and free electrons, have emerged as a preeminent platform for investigating light–matter interactions[8,9]. The advent of photon-induced near-field electron microscopy (PINEM) [10,11] has further propelled this field forward by providing femtosecond-resolved visualizations of electron–photon interactions[12,13]. When electron beams traverse SPP-enhanced optical field regions, the resultant EELS spectra exhibit multiple photon sidebands (see Fig. 1a and 1c), indicative of coherent multiphoton scattering[10,14,15]. These PINEM observations furnish direct evidence of laser control over quantum electron wavepackets and suggest novel routes toward quantum optics involving free electrons, photons, and surface plasmons[12,16–18].

More recently, there has been growing interest in extending plasmonic interactions from surfaces into the volume, where bulk plasmons (BPs) present themselves as promising candidates for next-generation quantum optical media [19–23]. In contrast to SPPs, which are inherently bound to interfaces and couple strongly to electromagnetic waves[24,25], BPs are longitudinal charge-density oscillations that propagate throughout conductors[26–28]. Their volumetric nature renders them attractive for subwavelength photonics, energy transport, and three-dimensional quantum architectures[23,29,30]. Nevertheless, the coherent optical excitation of BPs remains radically challenging [31,32]. First, the longitudinal polarization of BPs is orthogonal to the transverse electric field of light, preventing direct dipole coupling[31,33]. Second, typical BP excitation energies (>10 eV) [34] far exceed those of visible or near-infrared photons (1–3 eV)[35,36] (see Fig. 1), precluding efficient optical excitation by light. Although extreme nonlinear schemes- such as strong-field ionization[37] or multiphoton absorption by high-intensity laser- can, in principle, overcome this energy mismatch, they often induce deleterious thermal effect and material damages[38–41]. Consequently, the coherent, direct, and controllable light control of BP excitations remains effectively inaccessible[32].

To address this bottleneck, we shift focus on an indirect route that enables us transfer optical coherence to BPs via free electrons. Recent developments in PINEM have established that femtosecond laser fields can imprint both phase and energy information of structured electromagnetic fields onto quantum electron wavefunctions[10,12,42,43]. These photon-modulated electrons[44,45], can then undergo inelastic (or elastic) scattering within a thin bulk material to excite BPs. By employing this two-step scheme—"first light to electron, then electron to BPs"—we circumvent the intrinsic mode and energy mismatch between visible-light photon and BPs through the delivery of PINEM electrons, thereby formulating an indirect but viable pathway for coherently manipulating BPs by light.

Here, we theoretically proposed a full quantum electrodynamic framework that enables coherent coupling between light and bulk plasmons mediated by ultrafast free electrons, and then experimentally verified the coupling mechanism via PINEM measurements. The mechanism unfolds in two stages as we observed: optical coherence of incoming photons is initially encoded onto the electron wavefunction through PINEM process, and subsequently these photon-dressed electrons transfer coherence to BPs through inelastic electron scattering. This indirect pathway overcomes the mismatch between photons and BP, unlocking a previously inaccessible regime of nanoplasmonic engineering. Beyond experimentally demonstrating the feasibility of this indirect coupling, our framework predicts that the energy and phase information imprinted in the PINEM electron will manifest as distinct interference patterns in EELS observations. These interference fringes exhibit highly sensitive to phase modulation of photons, providing a new degree of freedom to manipulate collective excitations deep inside matter. Together, our findings not only establish a mechanism for coherent light–bulk plasmon interaction at femtosecond time scale but also open avenues toward free electron quantum optics and quantum nanoplasmonics beyond surface.

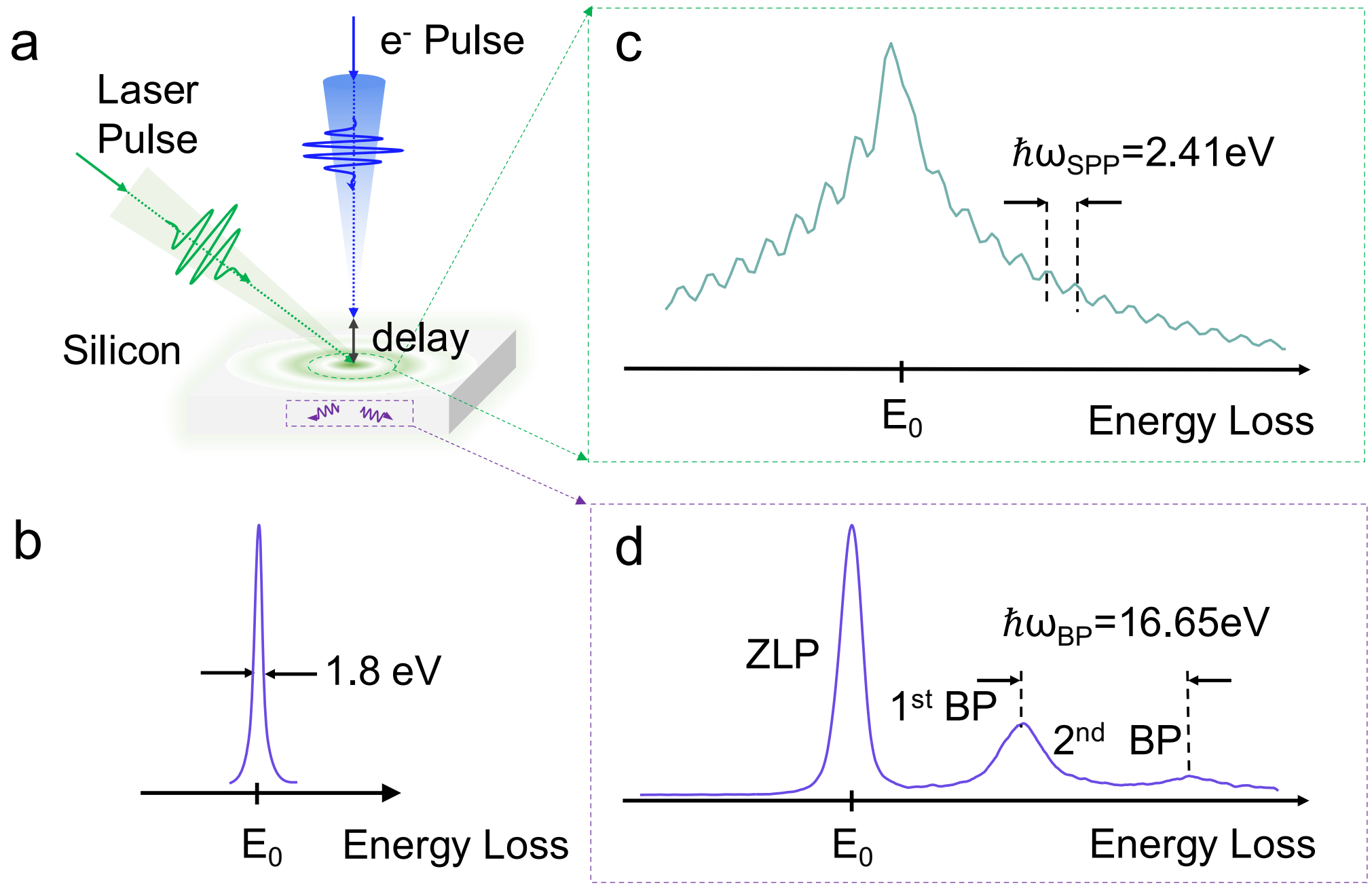

**Figure 1| Time-resolved free electron-light-matter interaction dynamics.** (a) Schematic of UTEM setup showing pulsed electrons and laser pulse interacting with a silicon sample under tunable time delay. (b) Initial electron pulses show the zero-loss peak (ZLP) spectrum at a central kinetic energy of 200 keV with a 1 eV FWHM. (c)

Electron spectrum of PINEM process, dressed by discrete photon sidebands spacing by $\hbar\omega_L$. (d) Electron spectrum of electron-BPs interaction for the spontaneous BPs excitation, showing the two loss peaks with equally spacing $\hbar\omega_{BP}$.

**QED Modeling.** Figure 2 depicts the UTEM setting alongside time-resolved EELS measurements, which inform our theoretical formulation. To capture the quantum mechanical underpinnings of indirect light–BP coupling mediated by free electrons, we develop a fully quantum electrodynamical (QED) framework that captures the coherent flow of information among photons, electrons, and BPs. As sketched in Fig. 2e, this mechanism is analogous to a Newton's cradle: incident photons (left) coherently transfer their phase and energy to ultrafast electrons, and these electrons subsequently couple to BPs (right) through inelastic scattering. Here, the electron–light interaction occurs via surface-bounded fields, we make no distinction between SPP and optical near-field. Crucially, akin to the middle balls in a Newton's cradle, the electron undergoes structured quantum modulation—its wavefunction acquires and retains coherent multiphoton features—underscoring its role as a coherent quantum mediator rather than a dissipative participant.

The interaction Hamiltonian $H_I$ comprises the electron-SPP coupling $H_{eSPP}$ and the electron-BP coupling $H_{eBP}$. The PINEM effect is induced by the coherent interaction of the electron with SPPs (details shown in Appendix B), which is derived from the minimal coupling $H_{eSPP} = -\frac{e}{m}\boldsymbol{A_{SPP}} \cdot \boldsymbol{p}$, we consider the near field with frequency $\omega_{SPP}$, $\hat{A}_{SPP} = \sum_q \frac{i\tilde{E}_z}{2\omega_q^-}\left[\hat{S}_q(t)e^{ik_q r} + \hat{S}_q^\dagger(t)e^{-ik_q r}\right]$, where $\hat{A}_{SPP}$, $\hat{S}_q(t)$ are electronmagnatic field operators, described by $H_{eSPP} = \sum_n [g_{eSPP} c_n^+ c_{n-1} S_1 + g_{eSPP}^* c_n^+ c_{n+1} S_1^+]$, where $g_{e-SPP} = -\frac{e\tilde{E}_z\hbar}{2\gamma m_e \omega^-} k_0$ is the effective coupling strength, $\tilde{E}_z$ denotes the longitudinal electric component driven by laser, $m_e$ the electron mass, $k_0 = p_0/\hbar$ the momentum of the incoming electron pulse and $\gamma$ the Lorentz factor (detailed derivations see Appendix B). Meanwhile, the electron-BP coupling $H_{eBP}$ originates from Coulomb scattering between electrons and BPs under dipole approximation, given by $H_{eBP} = -e\boldsymbol{d} \cdot \boldsymbol{E_c}$, where $-e\boldsymbol{d}$ denotes the atomic dipole moment and $\boldsymbol{E_c}$ is the Coulomb field of the electron. Decomposing $\boldsymbol{d}$ into bulk plasmon mode operators $B_q$, we have[19,46] $\boldsymbol{d} = \sum_q d_q \left(B_q^\dagger(t)e^{-i\boldsymbol{q}\cdot\boldsymbol{r}} + B_q(t)e^{i\boldsymbol{q}\cdot\boldsymbol{r}}\right)\hat{x}_q$ (see Appendix C). Here, $q$ denotes the wavenumber of a bulk plasmon mode, $\hat{x}_q$ is its polarization direction. While BPs form a quasi-continuous spectrum in $q$, only specific modes that satisfy the phase-matching condition with the electron contribute significantly to the coupling process. Upon second quantization (see Appendix D), the interaction becomes $H_{eBP} = \sum_n \left[ g_{e-BP} c_n^\dagger c_{n-1} B_1 + g^*{}_{e-BP} c_n^\dagger c_{n+1} B_1^\dagger \right]$ ,where $g_{e-BP} = -\frac{2e}{\gamma^2 R_0} d_1 \hat{R}_0 \cdot \hat{x}_1$ is the complex coupling strength.

Analogous to the leftmost ball in a Newton's cradle (Fig. 2e), the modulated electron ("the next ball") emerges from this stage bearing quantum information without dissipative loss and is ready to excite BPs in the subsequent interaction $H_{eBP}$. In this analogy, SPP photon coherently imprints its phase and energy onto the ultrafast electron via the PINEM interaction, preparing it as a quantum carrier of photon coherence [13], corresponding to the initial momentum transfer from the leftmost ball and culminates in the right-side energy transfer, thereby completing the coherent coupling pathway from SPPs to electrons, and ultimately to BPs, enabling the indirect communication between SPPs and BPs. To this point, all relevant degrees of freedom—the free electron, the photon, and the BP—are quantized. This unified QED treatment captures quantum coherence, correlations, and potential entanglement among the three subsystems, thereby transcending semiclassical theories of plasmonics. Quantizing each component further enables us to examine multiphoton scattering processes, electron-mediated coherent coupling, and the emergence of nonclassical entangled photon-BP states in a self-consistent manner.

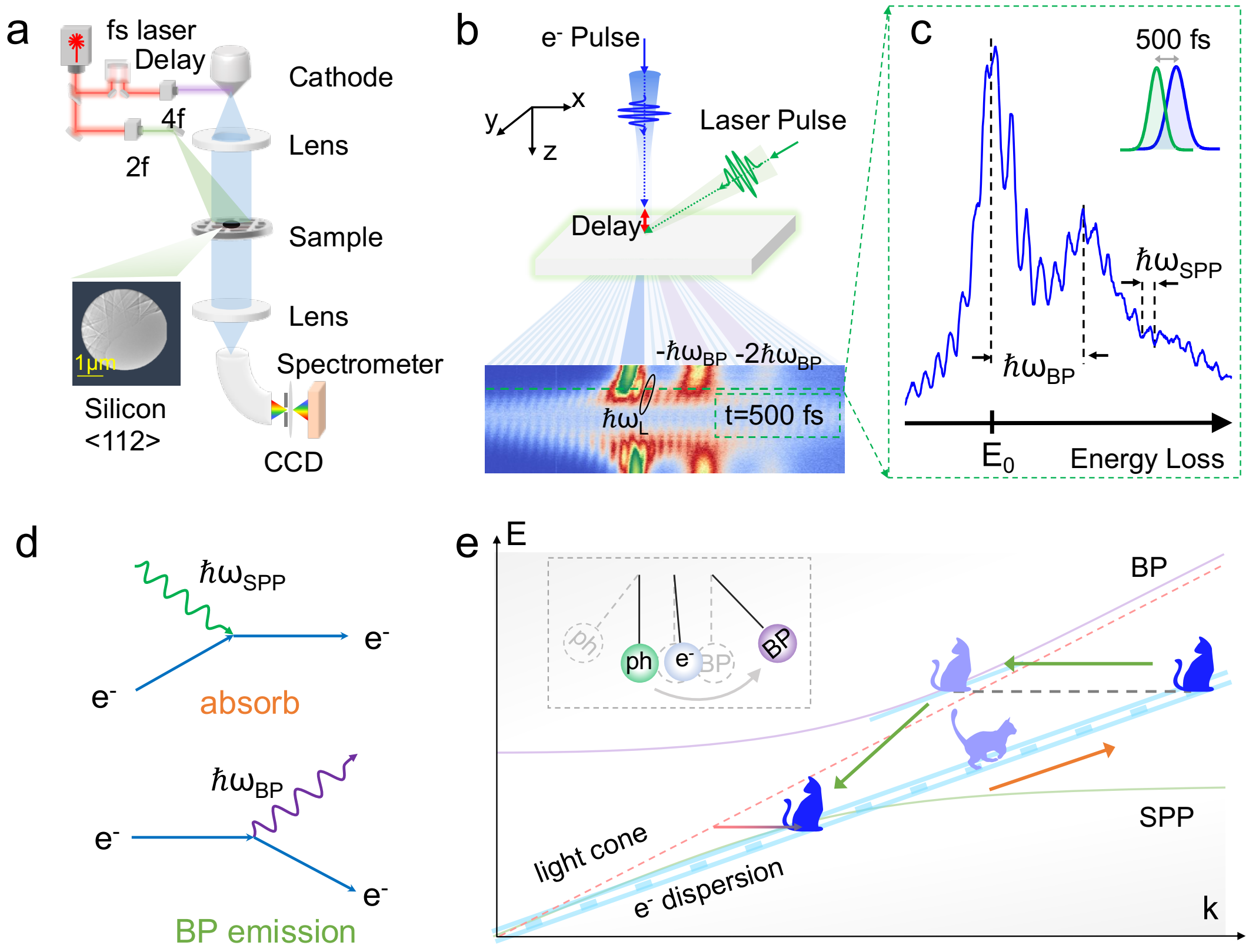

**Figure 2| UTEM setup for constructing indirect photon- BP coupling mediated by free electrons, through the subsequent photon-electron and electron-BP interaction.** (a) Schematic of the UTEM setup: a femtosecond laser is split into pump (near-field illumination) and probe (electron pulse excitation) arms via a beam splitter and delay stage; inset shows the TEM image of a silicon sample. (b) A 515 nm pump beam ($\hbar\omega$ = 2.41 eV) impinges to the surface, while the electron pulse transit the sample with a computer-controlled delay with the light beam, and the EELS scans shows both PINEM sidebands and BP loss peaks. (c) Slice of the EELS at 500 fs delay, depicts both multiphoton PINEM sidebands (2.41 eV spacing) and the BP loss peaks (16.65 eV). (d) The two fundamental electron scatterings, with BP excitation and photon absorption. (e) Two electron scattering channels: (i) direct multiphoton process, and (ii) mixed photon and BP process. Inset: schematic of indirect energy transfer from photon, to electron, and to BP, analogous to a classical Newton's cradle.

**Multiphoton versus bulk plasmon scattering**. Having established the theoretical framework, we now turn to the manifestation of this interaction in experiment—specifically, the co-occurrence of PINEM and BP scattering within the same electron–light–matter interaction event. Experimentally (Fig. 2a), free electrons passing through the sample region undergo inelastic energy changes, which we record via EELS detector, yielding a two-dimensional dataset (Fig. 2b): electron energy (horizontal axis, in eV) versus time delay between laser and electron pulses (vertical axis, in fs). A fixed-delay slice (Fig. 2c) reveals the energy spectrum at a given moment. Fig. 2e provides a schematic overview of the multiphoton scattering process through the plasmonic dispersion landscape. At a given energy, an electron absorbs multiple photons ($\hbar\omega_L = 2.41eV$), a process enhanced by near-field coupling with SPPs, and subsequently transfers this energy into a resonant BP mode ($\hbar\omega_{BP} = 16.65eV$). This sequence forms a coherent energy-transfer pathway from photons to BPs via the electron, reflecting a hybrid interaction process in which PINEM and BP scattering occur jointly rather than sequentially.

To describe how photon coherence is transferred to bulk plasmons, we treat the multiphoton interaction and BP excitation as a concurrent process, without assuming any strict time ordering. Although SPPs are confined to the surface and BPs reside deeper in the material, the femtosecond electron pulse is wide enough to interact with both within the same transit. As a result, the tail of the electron wavepacket can couple to SPPs while the front interacts with BPs, enabling spatial and temporal overlap. This spatiotemporal coherence justifies a unified quantum treatment for the multiphoton–electron-BP interaction, as shown in Fig. 2e. Prior to any interaction, the joint quantum state of the free electron, SPP mode, BP mode is a direct product: $|\psi_{\rm in}\rangle = |E_0\rangle \otimes$ $|\alpha\rangle_{SPP} \otimes |0\rangle_{BP}$, where $|E_0\rangle$ denotes the electron in its initial energy eigenstate at zero-loss energy ($E_0 = 200\text{keV}$), $|\alpha\rangle_{spp}$ is a coherent SPP state associated with the laser field, characterized by a mean plasmon occupation number $|\alpha|^2$, and the initial BP field is the vacuum state $|0\rangle_{BP}$, reflecting the absence of any excitation prior to the electron–field interaction. To explicitly track the energy transfer and phase evolution throughout the interaction, we expand the initial state in the Fock basis of the SPP mode:

$$|\psi_{\text{in}}\rangle = \sum_{j=0}^{\infty} e^{-\frac{|\alpha|^2}{2}} \frac{\alpha^j}{\sqrt{j!}} |E_0, j, 0\rangle = \sum_{j=0}^{\infty} c_j \, |E_0, j, 0\rangle. \tag{1}$$

This representation enables a clear quantum description of the full interaction process, facilitating our analysis of energy flow and information transfer across subsystems. The final state after interaction can be obtained via the scattering matrix $\hat{S}$, such that $|\psi_{\text{f}}\rangle = \hat{S}|\psi_{in}\rangle$. The scattering matrix encapsulates the concurrent quantum interaction between the electron, SPP, and BP, with assuming negligible electron dispersion in the sub relativistic regime ( $\beta = 0.7$ ), given by $\hat{S} = \hat{D}_{SPP}(g)\hat{D}_{BP}(g_{BP}) = \left(e^{g\hat{b}\hat{s}_1{}^{\dagger} - g^{*}\hat{b}^{\dagger}\hat{s}_1}\right)\left(e^{g_{BP}\hat{b}_{BP}\hat{B}_1{}^{\dagger} - g_{BP}^{*}\hat{b}_{BP}{}^{\dagger}\hat{B}_1}\right)$. Here, $g$ and $g_{BP}$ represent the respective coupling strengths for the SPP and BP interactions. The operators $\hat{b}$, $\hat{b}^{\dagger}$ and $\hat{b}_{BP}$, $\hat{b}_{BP}^{\dagger}$ are electron's ladder operator associated with the SPP and BP scattering channels, respectively, each satisfying bosonic-like commutation relations under the effective coupling model[47]: $[\hat{b}, \hat{b}^{\dagger}] = 0$ and $[\hat{b}_{BP}, \hat{b}_{BP}^{\dagger}] = 0$. Therefore, $\hat{S}$ acts as a composite displacement operator that[49] successively entangles the electron states with quantized SPP and then with BP modes, enabling coherent transfer between SPP and BP indirectly. The resulting final state after interaction is

$$|\psi_{\text{f}}\rangle = \sum_{j,n,n_{BP}} c_{j,n,n_{BP}} \left|E_{-n-7n_{BP}}, j+n, n_{BP}\right\rangle \tag{2}$$

where $E_{-n-7n_{BP}}$ indicates the net energy loss of the electron due to absorption or emission of n SPP modes and $n_{BP}$ BP modes. After some calculations (see Theoretical framework), the amplitude coefficients are given by

$$c_{j,n,n_{BP}} = \mathrm{e}^{-\frac{(|\alpha|^2 + |g_{BP}|^2)}{2}} \frac{\alpha^{j+n}}{\sqrt{(j+n)!}} e^{-in\phi_0} J_{-n}(2|G|) \frac{(g_{BP})^{n_{BP}}}{\sqrt{n_{BP}!}}. \tag{3}$$

Employing PINEM analysis under the assumption of a coherent initial SPP state, the final electron wavefunction emerges as a superposition of photon sidebands indexed by n, with amplitude $J_{-n}(2|G|)$[48]. The Bessel function argument $|G|$ is proportional to the laser amplitude $|\alpha|$ and the phase $\phi_0$ encodes the delay between the laser and electron pulse. In contrast, the subsequent e-BP interaction is treated perturbatively. Expanding to second order, the PINEM electron is further perturbed by one- and two-BP scattering contributions, yielding $\left|\psi_f^{(e)}\right\rangle = \sum_n J_{-n}(2|G|)\, e^{-in\phi_0}\left(|E_{-n}\rangle + \xi_1\left|E_{-n-n_{BP}}\right\rangle + \xi_2\left|E_{-n-2n_{BP}}\right\rangle\right)$, where $n_{BP} = \hbar\omega_{BP}/\hbar\omega_L = 7$ corresponds to the energy loss per BP excitation in silicon ($\hbar\omega_{BP} = 16.65\ eV$). The coefficients $\xi_1$ and $\xi_2$ quantify the contributions from single and two BP mode excitations. Notice that $n_{BP} > 0$, corresponds to no absorption of BP modes for the silicon material at equilibrium. These parameters, along with the multiphoton interaction strength $\eta_\omega, \xi_1, \xi_2$, can be extracted through numerical fitting of the experimental data. The complete dynamics, capturing both multiphoton and BP scattering from the electron's perspective- are illustrated in Fig. 3.

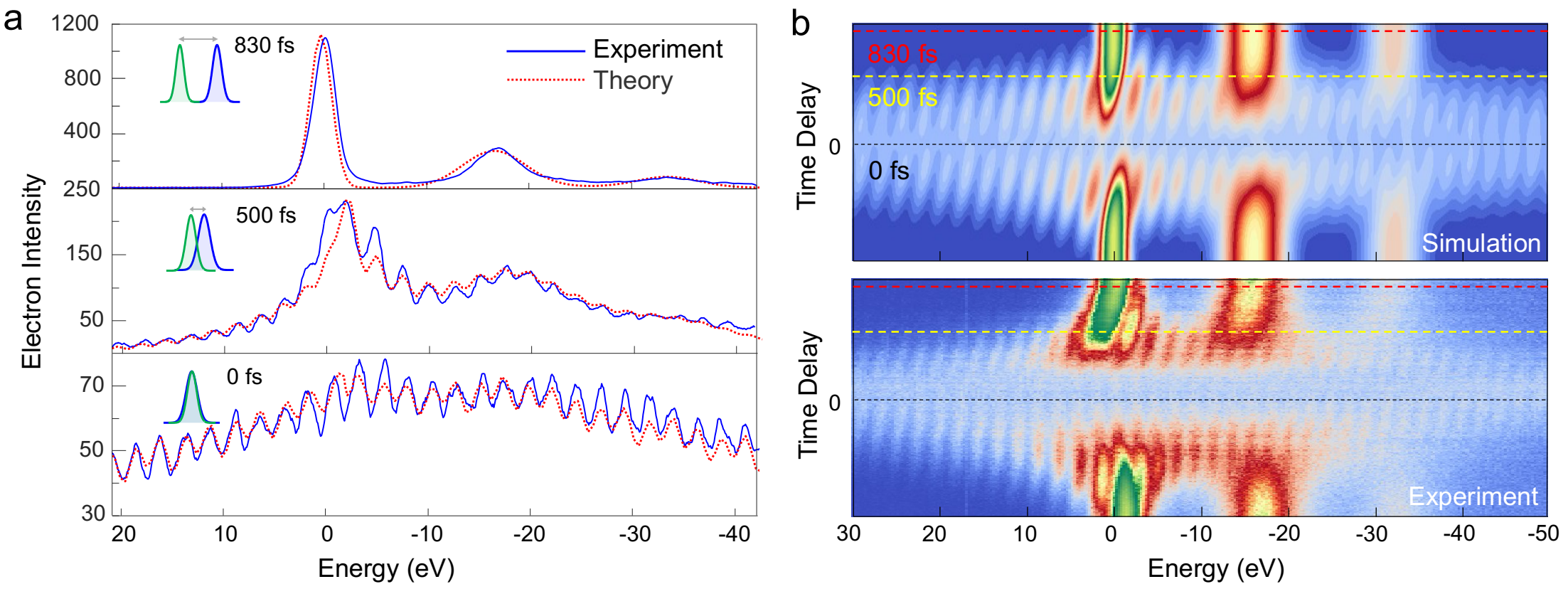

**Figure 3| Time-resolved EELS observation revealing the full electron–plasmon coupling dynamics.** (a) EELS spectra at different delays. At delay 830 fs, no multiphoton participated, only the associated bulk plasmon excitation losses appear. Af 500 fs, there are coexistence of bulk plasmon scattering and PINEM process. While at zero delay (0 fs), PINEM modulation dominates, almost suppressing the signature of BP scattering. (b) Numerical simulation of EELS spectrum in terms of delays, in excellent agreement with the experimental data.

**EELS measurements.** Figure 3 demonstrates the fitting process with the experimental data. The UTEM setup and sampling fabrication process are presented in Methods. Fig. 3b presents the time-delay-dependent EELS maps, both experimentally measured and numerically simulated. To elucidate the scattering dynamics, we highlight three representative regimes in Fig. 3a, respectively. At a delay of 830 fs, the laser pulse arrives before the electron, resulting in no temporal overlap. In this case, only a broad BP loss peak is observed. At delay of 500 fs, partial temporal overlap allows the electron to coherently interact with the optical nearfield (or SPP excitation), acquiring discrete energy sidebands. This electron then couples to the BP modes, resulting in the simultaneous observation of sidebands and BP loss features.

At zero delay, the optical and electron pulses coincide, maximizing interaction time. In this case, PINEM modulation dominates the EELS spectrum, while BP signatures become obscured suppressed by strong multiphoton scattering. This transition marks the full establishment of the photon–electron–bulk plasmon scattering sequence, with the electron indicating as a quantum transducer that imprints the photon phase and energy into the bulk. These measured spectrum at different delays are quantitatively captured by a semi-classical approximation that incorporates the full interaction Hamiltonian (see Method). The model accounts for the coherent nature of both PINEM and e–BP coupling and numerically fits with the observed sideband distributions and delay-dependent visibility with high accuracy.

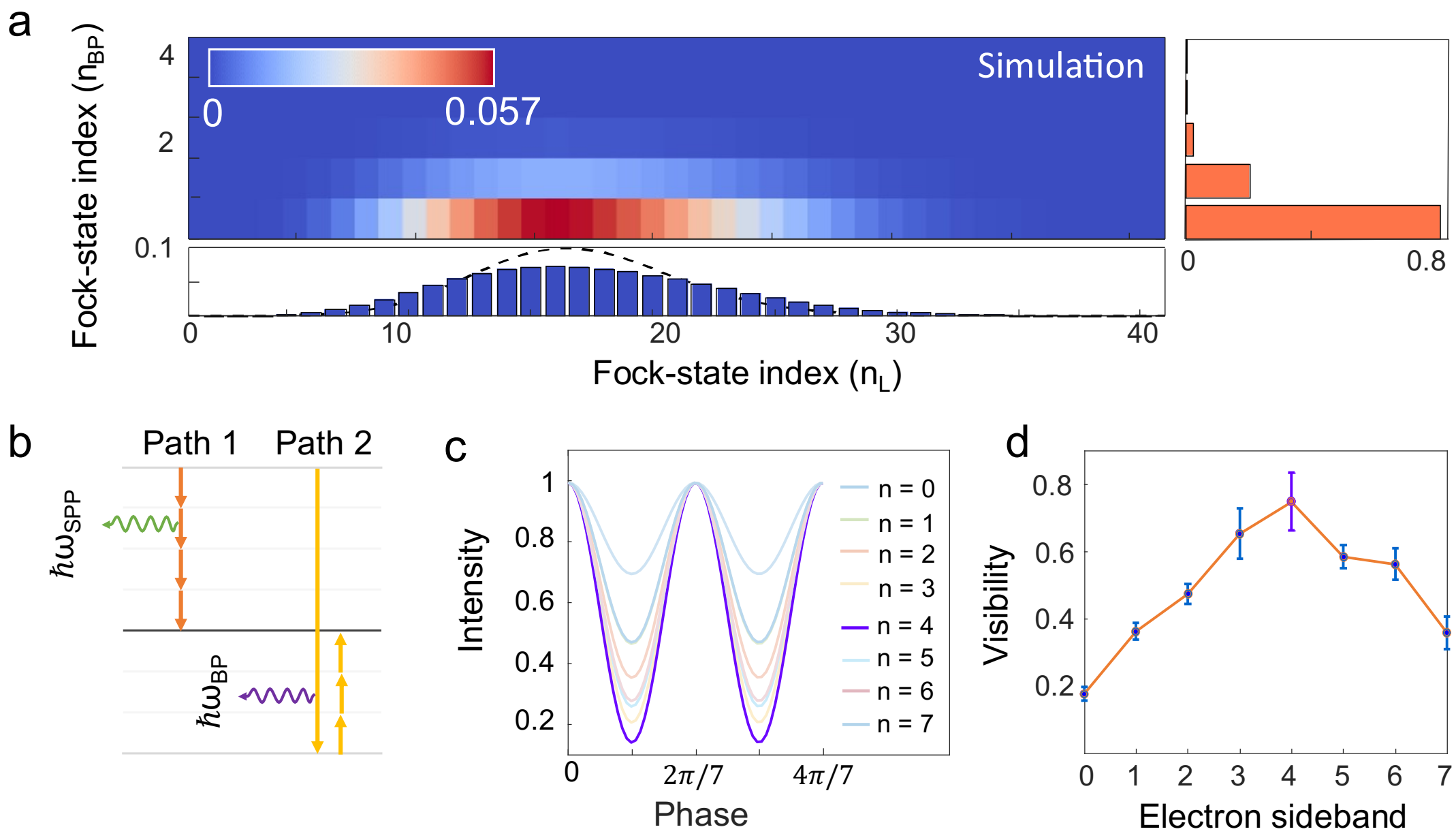

**Figure 4| Quantum emergence of spectral interference in the electron-photon–BP interaction.** (a) Joint Fock-state distribution of photon and BPs after tracing out electrons' degree of freedom.($g = 0.6$; $|\alpha| = 4$; $g_{BP} = 0.5$) (b) Two scattering pathways from sideband $E_0$ to $E_{-4}$, including photon emission (Path 1) and mixed photon-BP processes (Path 2). (c) The population of energy sideband $E_{-n}$ shows the two-channel interference in terms of the relative phase. (d) Maximum visibility is observed at sideband index $n = -4$, at delay 500 fs.

**Quantum regime of indirect coupling.** To explore the quantum regime of indirect photon–BP coupling, we examine the joint quantum state of photons and bulk plasmons after interaction, as shown in Fig. 4a. The horizontal and vertical axes correspond to SPPs and BPs Fock-state indices, respectively. The distribution reveals a clear correlation: photon number modulation occurs in tandem with the excitation of higher BP states, providing direct evidence of coherent energy sharing between the two subsystems mediated by the electron. This co-occurrence confirms the viability of indirect light–BP coupling via electron mediation and marks the onset of a quantum regime in which multiphoton and bulk plasmon processes coexist coherently. We consider an electron undergoes the scattering to the sideband level of $\mathrm{E}_{-4} = -4\hbar\omega_L$. This transition can be accessed through two quantum pathways: (i) direct emission of four photons; (ii) emission of a single bulk plasmon followed by absorption of three photons(Fig.4b). All two channels conserve energy and contribute coherently.

In what follows, we focus our on the two-channel interference, which suffices to capture the physics underlying the quantum coherent coupling mediated by an electron. The interference gives rise to a total final state as a superposition of two components: $\left|\psi_{\mathrm{f}}^{(1)}\right\rangle = \sum_{j,n} c_{j,n,0} \left|E_{-n}, j+n, 0\right\rangle$, with zero bulk plasmon excitation, and $\left|\psi_{\mathrm{f}}^{(2)}\right\rangle = \sum_{j,n'} c_{j',n',1} \left|E_{-n'-7}, j+n', 1\right\rangle$, involving a single excited BP mode ( $\mathrm{n_{BP}} = 1$ ). Therefore, the final state is given by:

$$|\psi_{\mathrm{f}}\rangle = \sum_{j,n} c_{j,n,0} \left|E_{-n}, j+n, 0\right\rangle + \sum_{j,n'} c_{j',n',1} \left|E_{-n'-7}, j+n', 1\right\rangle \qquad (4)$$

The coefficients $c_{j,n,0}$ and $c_{j',n',1}$ are given in Eq.3. To ensure quantum interference between the two paths, we require that the electron final energy be the same across both pathways, which yields the condition $n' - n = -7$. To probe the interference fringes associated the BP excitation, we project this final state onto a post-selected BP superposition state, namely $(|0\rangle + |1\rangle)/\sqrt{2}$. This yields an interference term in the -n-th sideband's population:

$$I_{-n} = \left(|J_{-n}(2|G|)|^2 + \left|J_{-(n-7)}(2|G|)\right|^2\right)/2 + |J_{-n}(2|G|)J_{-(n-7)}(2|G|)| \cos\left(7\phi_0 - (\phi_{\mathrm{BP}} - \phi_e)\right) \qquad (5)$$

where the expressions of Eq. 5 are used. Specially, the relative $\phi_0 = \phi_\omega - \phi_e$ is the difference between the laser pulse ($\phi_\omega$) and the electron phase ($\phi_e$), while $\phi_{BP}$ denotes the intrinsic phase of BP excitation. The first two terms in $I$ describe the independent contributions from the multiphoton-only and mixed photon–BP pathways, respectively. The third term captures their interference, governed by the relative phase $7\phi_0 - (\phi_{\mathrm{BP}} - \phi_e)$, involving three participating fields.

To quantitatively evaluate how phase modulation governs the re-distribution of electron sidebands, we introduce visibility as a sensitive measure for interference contrast. Defined as $\mathrm{V} = (I_{max} - I_{min})/(I_{min} + I_{max})$, this measure captures the relative modulation depth of each sideband weight under varying optical phase conditions. At a delay of 500 fs, the (-4)-th sideband (Fig. 4c, purple trace) exhibits the strongest modulation, corresponding to the highest visibility among all sideband orders (Fig. 4d). This high contrast arises from optimal spectral and temporal overlap between multiphoton absorption and bulk plasmon excitation, enabling coherent interference between scattering channels (Path 1 and Path 2). These observation suggest experimental evidence that quantum interference between light-driven and plasmon-mediated electron transitions can be resolved and selectively enhanced by femtosecond control. Notably, in our experiment the BP modes is excited spontaneously even with PINEM electrons, in which the phase $\phi_{\mathrm{BP}}$ is random, so that our EELS measurement cannot resolve the phase information of spontaneous BP excitation.

**Conclusion.** We have demonstrated a new mechanism for exciting bulk plasmons (BPs) indirectly – by employing ultrafast free electrons as coherent mediators of light-matter interaction. Through coherent photon-electron coupling, followed by subsequent electron–BP scattering, we realized a two-step coherent interaction that enables optical access to BPs without direct photon coupling. Our combined theoretical framework and time-resolved EELS measurements confirmed the co-occurrence of photon–electron and electron–BP interactions, indicating the role of the electron as an active intermediary facilitating coherent light–BP coupling. Beyond establishing this indirect multiphoton coupling, our findings open a route to control volumetric collective excitations -long overlooked in current nanoplasmonics. Furthermore, the demonstration of electron-mediated entangled photon-BP excitations paves the way for ultrafast imaging of bulk excitations and quantum plasmonic control, establishing a foundation for free-electron–based photonic technologies at nanoscale.

## Acknowledgements

Jun Li is supported by the Youth Innovation Promotion Association of CAS (2022004), the National Natural Science Foundation of China (Grant No. U22A6005), the Synergetic Extreme Condition User Facility (SECUF, https://cstr.cn/31123.02.SECUF).

Y. P. acknowledges the support of the NSFC (No. 2023X0201-417-03) and the fund of the ShanghaiTech University (Start-up funding).

The authors declare no competing financial interests.

Correspondence and requests for materials should be addressed to Y.P. (yiming.pan@shanghaitech.edu.cn), H.Y.(hxyang@iphy.ac.cn) and J.L.(junli@iphy.ac.cn).

## Methods

**UTEM Setup.** To experimentally validate the proposed electron-mediated light–BPs coupling, we implemented a PINEM experiment using an ultrafast transmission electron microscope (UTEM) as sketched in Fig. 2a, which was constructed by modifying a commercial dual aberration-corrected TEM (JEOL JEM-ARM200F NEOARM) equipped with Schottky-type field-emission gun and Gatan GIF Continuum K3 system. Ultrafast optical pulses (λ = 1030 nm, duration = 190 fs, 200 kHz repetition rate) were split into two synchronized arms. One arm was frequency-quadrupled (λ = 257 nm, $\hbar\omega$ = 4.82 eV) to photoemit femtosecond electron pulses from a $ZrO_2$-coated tungsten nanotip. The emitted electrons were accelerated to 200 keV and tightly focused onto a wedge-shaped silicon thin foil (<112> orientation), prepared via mechanical polishing, dimpling, and ion milling. The other optical arm was frequency-doubled (λ = 515 nm, $\hbar\omega$ = 2.41 eV) and directed obliquely (21.6° incidence angle) onto the sample surface with a 50 μm spot size (Fig. 2b). The temporal delay between optical and electron pulses was controlled with a motorized delay stage (resolution: 0.67 fs), enabling systematic mapping of ultrafast electron-light-BPs interactions across distinct time windows.

**Sample fabrication and EELS detection.** The silicon thin-foil specimen with <112> orientation for TEM observations was prepared by traditional "dimpling " method. The process begins by dicing a silicon wafer into a small (~5 mm) piece, which is then adhered to a flat metal support. The unmounted side is mechanically ground to approximately 100-150 μm thickness using coarse abrasives. The sample is flipped and mounted onto the support and further polished. Then the assembly undergoes dimpling: a specialized machine rotates the sample against polishing wheels while oscillating ti radially over progressively finer diamond abrasives (15 μm down to 1 μm), preferentially thinning the center of the silicon chip to create a concave dimple with a critical center thickness of 20-30 μm. The fragile dimpled chip is carefully released from the support, cleaned, and mounted onto a molybdenum TEM grid. The final step is argon ion milling at low energy (2-5 keV) and a shallow angle (3-6) with nitrogen cooling, to remove the damaged surface layer created by mechanical polishing and achieve electron transparency, typically until perforation occurs centrally within the dimple.

The electron energy gain/loss spectra were acquired by the post-column Gatan spectrometer with a 3456 × 3456 pixel direct electron detection camera with a typical exposure time of 1 s. The time-dependent PINEM single was collected with the aid of a homebrew Python script and constructed into two-dimensional data.

**Fitting and data analysis.** The simulation is carried out using a semiclassical model in which the electromagnetic field is treated classically (i.e., without second quantization), while the electron is fully quantum mechanical. All key parameters are chosen to closely match experimental conditions to ensure the relevance and accuracy of the modeling.

The electron wavepacket is modeled with a temporal duration of 200 fs (FWHM), and the laser pulse has a duration of 190 fs, consistent with measured experimental values. A chirp factor of 3 is applied to account for the temporal dispersion of the electron pulse. The initial energy distribution of the electron beam has a FWHM of 1.8 eV, and is broadened to 2.4 eV after accounting for interaction-induced broadening and instrumental effects.

Coupling parameters are selected based on experimental estimates and previous literature: the electron–photon coupling strength is set to $g = 30$, while the electron–bulk-plasmon coupling strength is taken as $g_{BP} = 0.5$. Due to the relatively long duration of the electron pulse, different segments of the wavepacket can interact with different fields at different times: while one part of the electron pulse is still interacting with the laser field at the sample surface (producing PINEM modulation), another part has already entered the material and is undergoing inelastic scattering with BPs.

The simulation accounts for this temporal overlap and spatial separation between the PINEM and BPs interaction regions. The resulting energy spectra are computed by propagating the electron wavefunction through the sequence of interactions, and the final simulated spectra are compared with experimental EELS data to extract key features and validate the presence of the indirect multiphoton coupling mechanism.

**Theoretical framework.** We formalize this picture with the time-dependent Schrödinger equation. The total Hamiltonian $H$ decomposes into a free Hamiltonian $H_0$ and an interaction part $H_I$ : $\mathrm{i}\hbar\frac{\partial}{\partial t}|\psi\rangle = H|\psi\rangle = (H_0 + H_I)|\psi\rangle$, where the composite wavefunction $|\psi\rangle = |\psi_e\rangle \otimes |\psi_{SPP}\rangle \otimes |\psi_{BP}\rangle$ comprises the free-electron state $|\psi_e\rangle$, $\mathrm{light-induced\ surface\ plasmon\ state}$ $|\psi_{SPP}\rangle$, and the bulk plasmon state $|\psi_{BP}\rangle$. And, the free Hamiltonian $H_0$ is given as follows $H_0 = \hbar\sum_n \varepsilon_n c_n^\dagger c_n + E_1^+ B_1^\dagger B_1 + E_1^- S_1^\dagger S_1$, includes the electron kinetic energy $H_e$, SPPs energy $H_{SPP}$ and the BPs energy $H_{BP}$. For analytical tractability, we assume that an electron pulse with negligible transverse divergence, thus neglecting dynamics perpendicular to the propagation direction, giving the 1D model. The eletron's kinetic Haminltonian is given by $H_e = \hbar\sum_n \varepsilon_n c_n^\dagger c_n$, where $c_n^\dagger$ ( $c_n$ ) generates (annihilates) a fast electron of energy $\hbar\varepsilon_n = \varepsilon_0 + m\hbar\omega_{SPP} + \hbar\omega_{BP}(n-m)/7(n = \pm1, \pm2, \cdots)$ , where $\varepsilon_0 = \sqrt{p_0^2c^2 + m^2c^4} = \gamma_0 mc^2$.

As illustrated in Fig. 2e, the interaction process can be viewed as a composite process: SPPs enable the coupling between the laser's longitudional photons $\omega_{spp}$ and the electron by providing the necessary synchronism condition; meanwhile, while BPs-with the phase-matched point in its dispersion curve at $(q, \omega_{BP}(q))$, allowing the quantum transitions between electron and BPs. The interactions enable effective multiphoton scattering between SPPs and BPs, mediated by the free electron. For our concern, we consider the single resonant bulk plasmon mode $B_1$ with $\hbar\omega_{BP}/\hbar\omega_{SPP} \approx 7$, as shown in Fig. 2c, we observe photon sidebands and BPs peaks. $H_{SPP} = E_1^+ B_1^\dagger B_1$ and $H_{BP} = E_1^- S_1^\dagger S_1$ (in Appendix A) are the quantized energy of SPPs and BPs, which is derived from the hybrid of light-plasmon polariton. In this hybrid basis, $E_1^\pm$ denote the upper and lower light-plasmon polariton branches, corresponding to SPPs and BPs excitations. To this point, we apply this Hamiltonian to solve the dynamics of the full system, obtaining the joint Fock-state distribution of photons and BPs after interaction with electrons.

Our primary motivation for developing this framework is to quantitatively investigate the coherent nature of the multiphoton–bulk plasmon coupling, as mediated by ultrafast electrons. In particular, this approach allows us to simulate the generation of electron energy sidebands and assess their phase-dependent visibility, which is characterized in Fig 4. We reveal the interference between distinct excitation pathways—such as purely multiphoton transitions and hybrid photon–BPs processes—and demonstrate that the phase coherence of the optical field can be imprinted and transferred to bulk plasmons.

We have the form of the displacement operator

$$\hat{D}(\hat{b}g) = e^{g\hat{b}\hat{s}^\dagger - g^*\hat{b}^\dagger\hat{s}} = e^{\frac{|g|^2}{2}} e^{-g^*\hat{b}^\dagger\hat{s}}\, e^{g\hat{b}\hat{s}^\dagger}$$

$$= e^{\frac{|g|^2}{2}} \sum_{m,l=0}^{\infty} \frac{(-g^*)^m (\hat{b}^\dagger)^m \hat{S}^m}{m!} \frac{g^l \hat{b}^l (\hat{S}^\dagger)^l}{l!}$$

Make a substitution

$$\mathrm{g} = -\mathrm{i}\frac{g_{e-spp}l}{\hbar v_e} = -\mathrm{g}^*$$

Similarly for $\widehat{D}(\hat{b}_{BP}g_{BP})$

$$\mathrm{g_{BP}} = -\mathrm{i}\frac{g_{e-BP}L}{\hbar v_e} = -\mathrm{g}_{\mathrm{BP}}^*$$

Both l and L are the interaction distance. Thus, we get the final state by $\hat{S}$.

$$|\psi_\mathrm{f}\rangle = \hat{S}|\psi_{in}\rangle$$
$$= \widehat{D}(\hat{b}g)\cdot\widehat{D}(\hat{b}_{BP}g_{BP})|\psi_{in}\rangle$$
$$= \mathrm{e}^{\frac{(-|\alpha|^2+|g|^2)}{2}}\widehat{D}(\hat{b}_{BP}g_{BP})\sum_{m,l,j=0}^{\infty}\frac{(-g^*)^m}{m!}\frac{g^l}{l!}\frac{\alpha^j}{\sqrt{j!}}\sqrt{\frac{(j+l)!}{(j+l-m)!}\frac{(j+l)!}{j!}}$$

$$|E_{m-l}, j+l-m, 0\rangle$$

define $l-m=n$

$$= \mathrm{e}^{\frac{(-|\alpha|^2+|g|^2)}{2}}\widehat{D}(\hat{b}_{BP}g_{BP})\sum_{m,l,j=0}^{\infty}\frac{(-|g|^2)^m g^n}{m!\,(m+n)!}\frac{(j+n+m)!}{j!\sqrt{(j+n)!}}\alpha^j|E_{-n}, j+n, 0\rangle$$
$$= \mathrm{e}^{\frac{(-|\alpha|^2+|g|^2+|g_{BP}|^2)}{2}}\sum_{m,j=0}^{\infty}\sum_{n=-j}^{\infty}\frac{(-|g|^2)^m g^n}{m!\,(m+n)!}\frac{(j+n+m)!}{j!\sqrt{(j+n)!}}\sum_{q,p=0}^{\infty}\frac{(g_{BP})^p(-g_{BP}^*)^q}{p!\,q!}$$

$$\frac{q!}{\sqrt{(q-p)!}}\alpha^j\left|E_{-n-7(q-p)}, j+n, q-p\right\rangle$$

define $q-p=n_{BP}$

$$= \mathrm{e}^{\frac{(-|\alpha|^2+|g|^2-|g_{BP}|^2)}{2}}\sum_{m,j=0}^{\infty}\sum_{n=-j}^{\infty}\frac{(-|g|^2)^m g^n}{m!\,(m+n)!}\frac{(j+n+m)!}{j!\sqrt{(j+n)!}}\alpha^j\sum_{n_{BP}=0}^{\infty}\frac{(g_{BP})^{n_{BP}}}{\sqrt{n_{BP}!}}$$

$$\left|E_{-n-7n_{BP}}, j+n, n_{BP}\right\rangle$$

In practical experiments, the parameter α is extremely large, approximately $10^9$. Due to the weak coupling between the light field and free electrons, α remains almost unchanged before and after the interaction. Ultimately only a very small amount of energy is exchanged between photons and electrons

$$|\mathrm{g}|^2 \ll 1$$

$$\langle j\rangle = |\alpha|^2$$

Finally, we use Bessel-function amplitudes to describe the final state,

$$|\psi_f\rangle = e^{-\frac{(|\alpha|^2+|g_{BP}|^2)}{2}} \sum_{n_{BP},j=0}^{\infty} \sum_{n=-j}^{\infty} \frac{\alpha^{j+n}}{\sqrt{(j+n)!}} e^{-in\phi_0} J_{-n}(2|G|) \frac{(g_{BP})^{n_{BP}}}{\sqrt{n_{BP}!}} \left|E_{-n-7n_{BP}}, j+n, n_{BP}\right\rangle$$

$$= \sum_{j,n,n_{BP}} c_{j,n,n_{BP}} \left|E_{-n-7n_{BP}}, j+n, n_{BP}\right\rangle$$

Here $G = g\sqrt{j} \approx g|\alpha|$ and $\phi_0 = \arg(\alpha) + \arg(-G^*) = \arg(\alpha G)$.

This methodology not only validates the proposed mechanism of indirect light–BP coupling via coherent electron mediation, but also establishes a robust platform for engineering and probing quantum interference phenomena in nanoplasmonics systems.

# Supplementary Material

# Indirect multiphoton scattering between light and bulk plasmons via ultrafast free electrons

Ruoyu Chen[1,2], Jun Li[3]*, Qiaofei Pan[4,5],Dingguo Zheng[3], Bin Zhang[6], Ye Tian[2]，Jianqi Li[3], Huaixin Yang[3]*, Yiming Pan[1]*

*[1]State Key Laboratory of Quantum Functional Materials, School of Physical Science and Technology and Center for Transformative Science, ShanghaiTech University, Shanghai 200031, China*

*[2] State Key Laboratory of Ultra-intense Laser Science and Technology, Shanghai Institute of Optics and Fine Mechanics, Chinese Academy of Sciences, Shanghai, China*

*[3]Beijing National Laboratory for Condensed Matter Physics, Institute of Physics, Chinese Academy of Sciences, Beijing, China*

*[4]Institute of Precision Optical Engineering, School of Physics Science and Engineering, Tongji University, Shanghai, 200092, China*

*[5]MOE Key Laboratory of advanced micro-structure materials, Shanghai, 200092, China*

*[6]School of Electrical Engineering - Physical Electronics, Center of Laser-Matter Interaction, Tel Aviv University, Ramat Aviv 69978, Israel*

## A Second Quantization of Plasmon

First, a 515 nm laser impinges on the sample surface, inducing photon-plasmon coupling[46] described by:

$$H_{PP} = \hbar\sum_{q} \omega_q a_q^\dagger a_q + \hbar\sum_{q} \tilde{\omega}_p b_q^\dagger b_q + \hbar\sum_{q} G_q(b_q^\dagger a_q + a_q^\dagger b_q) \qquad (A.1)$$

Here, $a_q(a_q^\dagger)$ is the photon creation (annihilation) operator of the incident laser for in-plane wave vector $q$, with dispersion $\omega_q = cq$. The plasmon creation (annihilation) operator $b_q$ ( $b_q^\dagger$ ) carries frequency $\tilde{\omega}_p = \omega_p - i\Gamma_p/2$ , where $\Gamma_p$ accounts for plasmon inelastic decay. The coupling constant $G_q$ quantifies the photon-plasmon interaction. The first two terms represent the free photon and plasmon Hamiltonians, the third term mediates their hybridization.

Using the basis set $a_n$ and $b_n$, we rewrite the Hamiltonian as follows and achieving its diagonalization through the application of the Bogoliubov transformation.

$$H_{PP} = \sum_{q} (a_q^\dagger \quad b_q^\dagger) \begin{bmatrix} \hbar\omega_q & \hbar G_q \\ \hbar {G_q}^\dagger & \hbar\tilde{\omega}_p \end{bmatrix} \begin{pmatrix} a_q \\ b_q \end{pmatrix} \qquad (A.2)$$

we let

$$A_q = \begin{bmatrix} \hbar\omega_q & \hbar G_q \\ \hbar {G_q}^\dagger & \hbar\tilde{\omega}_p \end{bmatrix} = U \begin{bmatrix} E_q^+ & 0 \\ 0 & E_q^- \end{bmatrix} U^\dagger = U h_{eff} U^\dagger \qquad (A.3)$$

The derived $E_q^-$ and $E_q^+$ represent the two branches of plasmon excitations. Specifically, $E_q^-$ corresponds to the surface plasmons polariton (SPP), whereas $E_q^+$ is associated with the bulk plasmons polariton. And we obtained the expression for the expansion coefficients as

$$u_{11} = \sqrt{\frac{|G_q|^2}{(\omega_p - \omega_q^+)^2 + |G_q|^2}} \qquad (A.4)$$

$$u_{21} = \sqrt{\frac{(\omega_p - \omega_q^+)^2}{(\omega_p - \omega_q^+)^2 + |G_q|^2}} \qquad (A.5)$$

$$u_{22} = \sqrt{\frac{|G_q|^2}{(\omega_p - \omega_q^-)^2 + |G_q|^2}} \tag{A.6}$$

$$u_{12} = \sqrt{\frac{(\omega_p - \omega_q^-)^2}{(\omega_p - \omega_q^-)^2 + |G_q|^2}} \tag{A.7}$$

The respective energy expressions for these branches are as follows:

$$E_k^{\pm} = \frac{\hbar}{2}(\omega_q + \widetilde{\omega}_p) \pm \frac{\hbar}{2}\sqrt{(\omega_q - \widetilde{\omega}_p)^2 + 4|G_q|^2} \tag{A.8}$$

When $cq = \omega_p$, the value of $|G_q|$ is determined as: $\omega_q^+ - \omega_q^- = 2|G_q|$.Thus, the Hamiltonian can now be expressed as:$H_{L-P} = \sum_q E_q^+ B_q^\dagger B_q + E_q^- S_q^\dagger S_q$

where

$$\begin{pmatrix} B_q \\ S_q \end{pmatrix} = U^\dagger \begin{pmatrix} a_q \\ b_q \end{pmatrix} \tag{A.9}$$

Since my laser emits monochromatic light and there is only one type of BPs in silicon, we set q=1 in this case.

$$H_{PP} = E_1^+ B_1^\dagger B_1 + E_1^- S_1^\dagger S_1 \tag{A.10}$$

Here, $\omega_{SPP} = \omega_q^-$ and $\omega_{BP} = \omega_q^+$

## B Electron–Surface Plasmon Interaction

The coupling between the electron and SPP is described by the following Hamiltonian:

$$H_{e-\mathrm{ph}} = -\frac{e}{\gamma m}\left[\vec{A}(z,t)\hat{p} + \hat{p}\vec{A}(z,t)\right] = -\frac{e}{\gamma m}\vec{A}(z,t)\hat{p} \tag{B.1}$$

In the Coulomb gauge, where $\nabla \cdot \vec{A} = 0$, $\gamma = \frac{1}{\sqrt{1-\beta^2}}$ is the Lorentz factor. The electric field represents the result of the coupling between light and SPP. Now, by performing second quantization on this electric field, we obtain:

$$\hat{A}(r,t) = \sum_q \frac{i\tilde{E}_z}{2\omega_q^-}\left[\hat{S}_q(t)e^{ik_q r} + \hat{S}_q^\dagger(t)e^{-ik_q r}\right] \tag{B.2}$$

This enables energy and momentum transfer to the electron. Using the Floquet-Bloch theorem, applicable to periodically driven systems, we express the electron's wave function $\psi_e(\mathbf{r},\mathrm{t}) = F_\perp(\boldsymbol{R})\sum_{\mathrm{n}}\mathrm{c_n}(t)\mathrm{e}^{(\mathrm{ik_n z})}$ . Here, $F_\perp(\boldsymbol{R})$ is the component perpendicular to the electron's incident direction. $\mathrm{R_0}$ is the impact parameter of the beam relative to the particle.

In this context, $\mathrm{k_n} = \frac{\mathrm{p_0}}{\hbar} + \frac{2\pi\mathrm{n}}{\Lambda}$ denotes the momentum component of the n-th harmonic. Typically, $\Lambda = \frac{2\pi}{\mathrm{q}}$ depends on the period of the intermediate medium (not the laser wavelength), while T is the laser's optical cycle. The amplitude $\mathrm{c_n}$ represents the slow-wave component of the n-th harmonic.

$$H_{e-ph} = -\frac{e}{\gamma m_e}\int dr^3 \hat{\psi}^\dagger(r,t)\hat{A}\cdot\hat{p}\hat{\psi}(r,t) \tag{B.3}$$

$$= -\frac{e}{\gamma m_e}\int dr^3 F_\perp(\boldsymbol{R})\sum_{\mathrm{n}} c_n^+(t)\mathrm{e}^{-(\mathrm{ik_n z})}\cdot\sum_q \frac{i\tilde{E}_z}{2\omega_q^-}\left[\hat{S}_q(t)e^{ik_q r} + \hat{S}_q^\dagger(t)e^{-ik_q r}\right]$$

$$\cdot\left(-i\hbar\frac{\partial}{\partial r}\right) F_\perp(\boldsymbol{R})\sum_{\mathrm{n'}} \mathrm{c_{n'}}(t)\mathrm{e}^{(\mathrm{ik_n z})}$$

We assume a well-collimated electron beam $|F_\perp(\boldsymbol{R})|^2 \approx \delta(\boldsymbol{R}-\boldsymbol{R}_0)$. $\mathrm{R_0}$ is the impact parameter of the beam relative to the particle. This way, the expression can be further simplified as follows:

$$= -\frac{e}{\gamma m_e}\int dr^3 \delta(\boldsymbol{R}-\boldsymbol{R}_0)\sum_{\mathrm{n}} c_n^+(t)\mathrm{e}^{-(\mathrm{ik_n z})}\cdot\sum_q \frac{i\tilde{E}_z}{2\omega_q^-}\left[\hat{S}_q(t)e^{ik_q r} + \hat{S}_q^\dagger(t)e^{-ik_q r}\right]$$

$$\cdot\left(-i\hbar\frac{\partial}{\partial r}\right)\delta(\boldsymbol{R}-\boldsymbol{R}_0)\sum_{\mathrm{n'}} \mathrm{c_{n'}}(t)\mathrm{e}^{(\mathrm{ik_{n'} z})}$$

$$= -\frac{e}{\gamma m_e}\int dr^3 |\delta(\boldsymbol{R}-\boldsymbol{R}_0)|^2 \sum_{\mathrm{nn'}q} \frac{\tilde{E}_z\hbar}{2\omega_q^-} c_n^+(t)\mathrm{c_{n'}}(t)\mathrm{e}^{-(\mathrm{ik_n z})}$$

$$\cdot\left[\hat{S}_q(t)e^{ik_q r} + \hat{S}_q^\dagger(t)e^{-ik_q r}\right]\cdot\left(\frac{\partial}{\partial r}\right)\mathrm{e}^{(\mathrm{ik_{n'} z})}$$

The constant terms are absorbed into the operators, resulting in:

$$e^{ikr} = e^{ik_{qx}x + ik_{qy}y + ik_{qz}z} = e^{ik_{qR_0}R_0 + ik_{qz}z} \tag{B.4}$$

$$\hat{S}'_q(t) = \hat{S}_q(t)e^{ik_{qR_0}R_0} \tag{B.5}$$

$$\hat{S}_q^{\dagger\prime}(t) = \hat{S}_q^\dagger(t)e^{-ik_{qR_0}R_0} \tag{B.6}$$

Then we get

$$=\sum_q -\frac{e\tilde{E}_z\hbar}{2\gamma m_e\omega_q^-}\int dz\sum_{\mathrm{nn'}} c_n^+(t)\mathrm{c_{n'}}(t)\mathrm{e}^{-\mathrm{i}(\mathrm{k_n-k_{n'}})}\cdot\left[\hat{S}_q{}'(t)e^{ik_{qz}z}+\hat{S}_q^{\dagger}{}'(t)e^{-ik_{qz}z}\right]k_{n'}$$

$$=\sum_q -\frac{e\tilde{E}_z\hbar}{2\gamma m_e\omega_q^-}\int dz\sum_{\mathrm{nn'}} k_{n'}c_n^+(t)\mathrm{c_{n'}}(t)\left[\hat{S}_q{}'(t)e^{i(\mathrm{k_n-k_{n'}}+k_{qz})z}+\hat{S}_q^{\dagger}{}'(t)e^{-i(\mathrm{k_n-k_{n'}}-k_{qz})z}\right]$$

with the rotation wave approximation, the phase matching condition is satisfied: $\mathrm{k_n}-\mathrm{k_{n'}}+k_{q_z}=\mathrm{k_n}-\mathrm{k_{n'}}-k_{q_z}=0$. This approach elegantly simplifies the integral:

$$=\sum_{nq} -\frac{e\tilde{E}_z\hbar}{2\gamma m_e\omega_q^-}k_0\left[c_n^+c_{n-q}S_q+c_n^+c_{n+q}S_q^+\right]$$

Since the incident laser is monochromatic and has a unique incident angle, the excited $q$ has only one mode. Thus, our equation finally simplifies to:

$$=\sum_n\left[g_{e-SPP}c_n^+c_{n-1}S_1+g_{e-SPP}^*c_n^+c_{n+1}S_1^+\right] \quad (\mathrm{B.7})$$

Where $g_{e-SPP}=-\frac{e\tilde{E}_z\hbar}{2\gamma m_e\omega^-}k_0$. Since $\mathrm{q}\ll k_0$, we approximate $\mathrm{k}_{n-q}\simeq\mathrm{k}_{n+q}\simeq\mathrm{k_0}$. where $g_{e-SPP}=-\frac{e\tilde{E}_z\hbar}{2\gamma m_e\omega^-}k_0$ is the effective coupling constant between the electron and the SPP field, where $\tilde{E}_z$ denotes the SPP's longitudinal electric field amplitude driven by laser. This Hamiltonian encodes SPP-mediated PINEM transitions, where the term $g_{eSPP}c_n^+c_{n-1}S_1$ corresponds to the absorption of an SPP quantum by the electron, inducing an upward energy step on the synthetic ladder, while the conjugate term represents the process of stimulated emission and downward transition.

## C Second Quantization of Dipole

The interaction between the electron and BPs is described within dipole approximation. The dipole moment $-e\boldsymbol{d}$ is determined by the position operator $d\hat{\boldsymbol{x}}$ via

$$-e\boldsymbol{d}=-ed\hat{\boldsymbol{x}} \quad (\mathrm{C.1})$$

We assume an atom with only two energy eigenvalues is described by a two-dimensional state space spanned by the two energy eigenstates $|e\rangle$ and $|g\rangle$. So we can write down an arbitrary state $|\psi\rangle=c_g|g\rangle+c_e|e\rangle$. Then the expectation value for the dipole moment of an atom in this state is

$$\langle\psi|\boldsymbol{d}|\psi\rangle=d\left(|c_e|^2\langle e|\hat{\boldsymbol{x}}|e\rangle+c_ec_g^*\langle g|\hat{\boldsymbol{x}}|e\rangle+c_e^*c_g\langle e|\hat{\boldsymbol{x}}|g\rangle+\left|c_g\right|^2\langle g|\hat{\boldsymbol{x}}|g\rangle\right) \quad (\mathrm{C.2})$$

The atoms posses inversion symmetry, therefore, energy eigenstates must be symmetric or anti-symmetric, i.e. $\langle e|\hat{\boldsymbol{x}}|e\rangle = \langle g|\hat{\boldsymbol{x}}|g\rangle = 0$. We obtain

$$\langle\psi|\boldsymbol{d}|\psi\rangle = d\big(c_e c_g^*\langle g|\hat{\boldsymbol{x}}|e\rangle + c_e^* c_g\langle e|\hat{\boldsymbol{x}}|g\rangle\big). \qquad \text{(C.3)}$$

In this equation, fiert term describes a transition fron $|e\rangle$ to $|g\rangle$, corresponding to energy emission; the other term represents the reverse process of the first. Accordingly, we describe this process using the ladder operators of the quantized BP field.

$$\boldsymbol{d} = \sum_q d_q\left(B_q^\dagger(t)e^{-iq\cdot r} + B_q(t)e^{iq\cdot r}\right)\hat{x}_q \qquad \text{(C.4)}$$

Here, we use $q$ as the index to denote the various BPs mode.

## D Electron–Bulk Plasmon Interaction

In the electron's own reference frame, the Coulomb field at position $\boldsymbol{r}'$ simplifies to: $\boldsymbol{E}_c'(\boldsymbol{r}') = e\nabla'\frac{1}{r'}$. Given that 200 keV electrons propagates at $\mathrm{v} \approx 0.7\mathrm{c}$ in transmission electron microscopy, a relativistic time-dilation correction is applied. By applying a coordinate transformation with $\boldsymbol{R}' = -\boldsymbol{R}$, $\mathrm{z}' = -\gamma\mathrm{z}$, $\boldsymbol{E_R} = \gamma\boldsymbol{E_{R'}}$ and $\mathrm{E_Z} = E_z'$, we obtain:

$$\boldsymbol{E_c} = -e\left(\gamma\nabla_{\mathrm{R}},\frac{1}{\gamma}\partial_z\right)\frac{1}{|\boldsymbol{R},\gamma z|} = \mathrm{e}\gamma\,\frac{\boldsymbol{R}+\boldsymbol{z}}{\left(\sqrt{x^2+y^2+(\gamma z)^2}\right)^3} \qquad \text{(D.1)}$$

Here, our electronic wave function is $\psi_e(\mathbf{r},\mathrm{t}) = F_\perp(\boldsymbol{R})\sum_{\mathrm{n}}\mathrm{c_n}(\mathrm{t})\mathrm{e}^{\mathrm{i}k_{\mathrm{n}}\mathrm{z}}$. In a uniform cylindrical charge density moving at ultra-relativistic speed, the transverse space charge forces disappear. This is due to the cancellation of electric and magnetic forces, which causes the transverse force to vanish at a rate proportional to $1/\gamma^2$.

$$H_{eBP} = -\int dr^3\hat{\psi}^\dagger(r,t)e\boldsymbol{d}\cdot\frac{1}{\gamma^2}\boldsymbol{E}_c'(\boldsymbol{r}')\hat{\psi}(r,t) \qquad \text{(D.2)}$$

$$= -\int dr^3F_\perp(\boldsymbol{R})\sum_{\mathrm{n}}\mathrm{c_n^\dagger}(\mathrm{t})\mathrm{e}^{-\mathrm{i}k_{\mathrm{n}}\mathrm{z}}\sum_q d_q\left(B_q^\dagger(t)e^{-iq\cdot r} + B_q(t)e^{iq\cdot r}\right)\hat{x}_q$$

$$\cdot\frac{e}{\gamma^2}\frac{\boldsymbol{R}+\boldsymbol{z}}{\left(\sqrt{x^2+y^2+z^2}\right)^3}F_\perp(\boldsymbol{R})\sum_{\mathrm{m}}\mathrm{c_m}(\mathrm{t})\mathrm{e}^{\mathrm{i}k_{\mathrm{m}}\mathrm{z}}$$

$$= -\frac{e}{\gamma^2}\sum_{nmq} \mathrm{c}_\mathrm{n}^\dagger(\mathrm{t})\,\mathrm{c}_\mathrm{m}(\mathrm{t}) d_q \int d\boldsymbol{R}\ |F_\perp(\boldsymbol{R})|^2 \hat{x}_q \int \mathrm{d}\boldsymbol{z}\ \frac{\boldsymbol{R}+\boldsymbol{z}}{\left(\sqrt{x^2+y^2+z^2}\right)^3}\Big(B_q^\dagger(t)e^{-iq_z z}e^{-i(q_x x+q_y y)}$$

$$+ B_q(t)e^{ik_q z}e^{-i(q_x x+q_y y)}\Big)\mathrm{e}^{-\mathrm{ik_n z}}\mathrm{e}^{\mathrm{ik_m z}}$$

Here, we approximate the integral $|F_\perp(\boldsymbol{R})|^2 \approx \delta^2(\boldsymbol{R}-\boldsymbol{R}_0)$ , which simplifies the entire equation

$$= -\frac{e}{\gamma^2}\sum_{nmq} \mathrm{c}_\mathrm{n}^\dagger(\mathrm{t})\,\mathrm{c}_\mathrm{m}(\mathrm{t}) d_q \int d\boldsymbol{R}\ \delta^2(\boldsymbol{R}-\boldsymbol{R}_0)\hat{x}_q$$

$$\cdot \int \mathrm{d}\boldsymbol{z}\ \frac{\boldsymbol{R}+\boldsymbol{z}}{\left(\sqrt{x^2+y^2+z^2}\right)^3}\Big(B_q^\dagger(t)e^{-iq_z z}e^{-i(q_x x+q_y y)}$$

$$+ B_q(t)e^{ik_q z}e^{-i(q_x x+q_y y)}\Big)\mathrm{e}^{-\mathrm{ik_n z}}\mathrm{e}^{\mathrm{ik_m z}}$$

$$= -\frac{e}{\gamma^2}\sum_{nmq} \mathrm{c}_\mathrm{n}^\dagger(\mathrm{t})\,\mathrm{c}_\mathrm{m}(\mathrm{t}) d_q \int \mathrm{d}\boldsymbol{z}\ \frac{\boldsymbol{R_0}+\boldsymbol{z}}{\left(\sqrt{R_0^2+z^2}\right)^3}$$

$$\cdot \hat{x}_q\left(B_q^\dagger(t)e^{-iq_z z}e^{-i(q_{x_0}x_0+q_{y_0}y_0)}\right.$$

$$\left.+ B_q(t)e^{ik_q z}e^{-i(q_{x_0}x_0+q_{y_0}y_0)}\right)\mathrm{e}^{-\mathrm{ik_n z}}\mathrm{e}^{\mathrm{ik_m z}}$$

Now, we can express the equation in a simpler form:

$$\tilde{B}_q^\dagger(\mathrm{t}) = B_q^\dagger(t)e^{-i(q_{x_0}x_0+q_{y_0}y_0)} \tag{D.3}$$

$$\tilde{B}_q(t) = B_q(t)e^{-i(q_{x_0}x_0+q_{y_0}y_0)} \tag{D.4}$$

Then we have

$$= -\frac{e}{\gamma^2}\sum_{nmq} \mathrm{c}_\mathrm{n}^\dagger(\mathrm{t})\,\mathrm{c}_\mathrm{m}(\mathrm{t}) d_q \int \mathrm{d}\boldsymbol{z}\ \frac{\boldsymbol{R_0}+\boldsymbol{z}}{\left(\sqrt{R_0^2+z^2}\right)^3}$$

$$\cdot \hat{x}_q\left(\tilde{B}_q^\dagger(\mathrm{t})e^{-iq_z z} + \tilde{B}_q^\dagger(\mathrm{t})e^{ik_q z}\right)\mathrm{e}^{-\mathrm{ik_n z}}\mathrm{e}^{\mathrm{ik_m z}}$$

At this point, the phase matching condition must also be satisfied, which is $\mathrm{k_m} - k_n - q_z = 0, \mathrm{k_m} - k_n + q_z = 0$.

$$= -\frac{e}{\gamma^2}\sum_{nmq} \mathrm{c}_\mathrm{n}^\dagger(\mathrm{t})\,\mathrm{c}_\mathrm{m}(\mathrm{t}) d_q \int \mathrm{d}\boldsymbol{z}\ \frac{\boldsymbol{R_0}+\boldsymbol{z}}{\left(\sqrt{R_0^2+z^2}\right)^3}$$

$$\cdot \hat{x}_q\left(\tilde{B}_q^\dagger(\mathrm{t})\delta(\mathrm{k_m} - k_n - q_z) + \tilde{B}_q^\dagger(\mathrm{t})(\mathrm{k_m} - k_n + q_z)\right)$$

$$= -\frac{e}{\gamma^2}\sum_{nmq} \mathrm{c}_\mathrm{n}^\dagger(\mathrm{t})\,\mathrm{c}_\mathrm{m}(\mathrm{t}) d_q \int \mathrm{d}\mathbf{z} \left[\frac{R_0}{\left(\sqrt{R_0^2+z^2}\right)^3}\hat{R}_0\cdot\hat{x}_q + \frac{z}{\left(\sqrt{R_0^2+z^2}\right)^3}\hat{z}\right.$$

$$\left.\cdot\,\hat{x}_q\right]\left(\tilde{B}_q^\dagger(\mathrm{t})\delta(\mathrm{k_m}-k_n-q_z)+\tilde{B}_q^\dagger(\mathrm{t})(\mathrm{k_m}-k_n+q_z)\right)$$

$$= -\frac{e}{\gamma^2 R_0}\sum_{nmq} \mathrm{c}_\mathrm{n}^\dagger(\mathrm{t})\,\mathrm{c}_\mathrm{m}(\mathrm{t}) d_q \left[\frac{z}{\sqrt{R_0^2+z^2}}\hat{R}_0\cdot\hat{x}_q - \frac{1}{\sqrt{R_0^2+z^2}}\hat{z}\cdot\hat{x}_q\right]_{-\infty}^{\infty}\left(\tilde{B}_q^\dagger(\mathrm{t})\delta(\mathrm{k_m}\right.$$

$$\left.-\,k_n-q_z)+\tilde{B}_q^\dagger(\mathrm{t})(\mathrm{k_m}-k_n+q_z)\right)$$

$$= -\frac{e}{\gamma^2 R_0}\sum_{nmq} \mathrm{c}_\mathrm{n}^\dagger(\mathrm{t})\,\mathrm{c}_\mathrm{m}(\mathrm{t}) d_q 2\hat{R}_0$$

$$\cdot\,\hat{x}_q\left(\tilde{B}_q^\dagger(\mathrm{t})\delta(\mathrm{k_m}-k_n-q_z)+\tilde{B}_q^\dagger(\mathrm{t})(\mathrm{k_m}-k_n+q_z)\right)$$

$$= -\frac{2e}{\gamma^2 R_0}\sum_{nq}(\tilde{B}_q^\dagger(\mathrm{t})\mathrm{c}_\mathrm{n}^\dagger(\mathrm{t})\,\mathrm{c}_\mathrm{n+q}(\mathrm{t})d_q + \tilde{B}_q^\dagger(\mathrm{t})\mathrm{c}_\mathrm{n}^\dagger(\mathrm{t})\mathrm{c}_\mathrm{n-q}(\mathrm{t})d_q)\hat{R}_0\cdot\hat{x}_q$$

$$= \sum_{nq} -\frac{2e}{\gamma^2 R_0} d_q(\tilde{B}_q^\dagger(\mathrm{t})\mathrm{c}_\mathrm{n}^\dagger(\mathrm{t})\,\mathrm{c}_\mathrm{n+q}(\mathrm{t}) + \tilde{B}_q^\dagger(\mathrm{t})\mathrm{c}_\mathrm{n}^\dagger(\mathrm{t})\mathrm{c}_\mathrm{n-q}(\mathrm{t})d_q)\hat{R}_0\cdot\hat{x}_q$$

$$= \textstyle\sum_{n,q}\left[g_{e-BP}c_n^+c_{n-q}B_q + {g^*}_{e-BP}c_n^+c_{n+q}B_q^\dagger\right]$$

That's means:

$$H_{eBP} = \sum_{n,q}\left[g_{e-BP}c_n^+c_{n-q}B_q + {g^*}_{e-BP}c_n^+c_{n+q}B_q^\dagger\right] \qquad (\mathrm{D.5})$$

where $g_{e-BP} = -\frac{2e}{\gamma^2 R_0} d_q\hat{R}_0\cdot\hat{x}_q$ is complex coupling constants.Since the incident laser is monochromatic and has a unique incident angle, the excited $q$ has only one mode. Thus, our equation finally simplifies to:

$$H_{eBP} = \sum_{n}\left[g_{e-BP}c_n^\dagger c_{n-1}B_1 + {g^*}_{e-BP}c_n^\dagger c_{n+1}B_1^\dagger\right] \qquad (\mathrm{D.6})$$

where $g_{e-BP} = -\frac{2e}{\gamma^2 R_0} d_1\hat{R}_0\cdot\hat{x}_1$ is the complex coupling strength between an electron sideband transition. The first term corresponds to the electron losing energy $\hbar\omega_{BP}$ to excite a BP(annihilating one electron quantum and creating a BP excitation), while the second term describes the reverse process: the electron gains energy $\hbar\omega_{BP}$ via stimulated emission from an existing BP. Together, these terms account for coherent bidirectional exchange of energy between the electron and BP field, essential for mediating photon–BP coupling via the electron's modulated quantum state.